# بررسی تأثیر غلظت عامل احیاکننده و pH بر خصوصیات ساختاری نانوذرات مغناطیسی مگنتایت جانشانی شده با روی

بهاره رضایی[1]*، احمد کرمانپور[2]، شیدا لباف[3]

## چکیده

امروزه نانوذرات فریت‌های اسپینلی به دلیل ویژگی‌های برتر مغناطیسی و زیست سازگار بودن، در عرصه پزشکی برای درمان سرطان، رهایش دارو، و تصویربرداری تشدید مغناطیسی کاربرد گسترده‌ای یافته‌اند. به دلیل وابستگی شدید خواص مغناطیسی به مورفولوژی و سایر جنبه‌های ساختاری، خواص مغناطیسی این نانوذرات به فرایند ساخت آن‌ها وابسته است. یکی از اهداف محققان بهبود ویژگی‌های مغناطیسی این نانوذرات از طریق جانشانی عناصری از قبیل منگنز، کبالت، روی، نیکل، منیزیم و آهن است. در این پژوهش، ابتدا نانوذرات مغناطیسی $Zn_{0.3}Fe_{0.6}O_4$ به روش هیدروترمال در حضور اسید سیتریک به عنوان عامل احیاکننده سنتز شد. برای کاهش میزان ناخالصی و حصول فاز اسپینل خالص و کاهش اندازه نانوذرات مغناطیسی، فرایند هیدروترمال در حضور مقادیر مختلف غلظت عامل احیا و pH انجام شد. بررسی‌های ساختاری نانوذرات حاصل با آزمون‌های پراش پرتو ایکس (XRD) و میکروسکوپ الکترونی روبشی (SEM) انجام شد. نتایج نشان داد که برای حصول نانوذرات اسپینلی خالص و تکفاز با اندازه و توزیع اندازه ذرات مناسب، کنترل غلظت عامل احیاکننده و pH ضروری است.

1- دانشجوی کارشناسی ارشد، شناسایی و انتخاب مواد، دانشکده مهندسی مواد، دانشگاه صنعتی اصفهان، اصفهان 84156-83111، ایران
bahar.rezaei@ma.iut.ac.ir
2- استاد، نانو مواد و شبیه سازی فرآیند، دانشکده مهندسی مواد، دانشگاه صنعتی اصفهان، اصفهان 84156-83111، ایران
3- استادیار، بیومواد، دانشکده مهندسی مواد، دانشگاه صنعتی اصفهان، اصفهان 84156-83111، ایران

کلمات کلیدی: کلمات کلیدی: نانوذرات مغناطیسی، فریت روی، روش هیدروترمال، عامل احیا کننده

**مقـدمـه**

نانو ذرات مغناطیسی می‌توانند کاربردهای بالقوه زیادی در زمینه‌های مختلف مانند ابزارهای ذخیره اطلاعات، عکس برداری از بدن، ذخیره‌سازی انرژی، ذخیره‌سازی اطلاعات مغناطیسی زیست پزشکی مانند زیست حسگرها، رهایش دارو، به عنوان عوامل کنتراست برای تصویربرداری رزونانس مغناطیسی (MRI) و فراگرمایی برای درمان سرطان دارند [2].

برای آن که نانو ذرات مغناطیسی بتوانند کاربردهای زیستی داشته باشند، باید بین 10-100 نانومتر باشد؛ چون در اندازه‌ی کمتر از ده نانومتر از طریق کلیه ها دفع می‌شوند و در مقیاس بزرگتر از 200 نانومتر به عنوان یک جسم خارجی، به وسیله‌ی سیستم دفاعی بدن از بین می روند. ویژگی مغناطیسی ذرات به اندازه، شکل، نسبت اجزای سازنده (برای مثال نوع نمک استفاده شده برای تهیه هسته مغناطیسی، نسبت بین ذرات مغناطیسی، pH و قدرت یونی محیط) و پوشش اطراف نانوذرات بستگی دارد[3]. نانوذرات مغناطیسی که در زیست پزشکی مورد بررسی قرار گرفته اند، بیشتر شامل اکسید-های آهن هستند [4]. از میان اکسیدهای آهن نانوذرات مگنتایت($Fe_3O_4$) یکی از گزینه‌های مهم مورد استفاده در زمینه پزشکی است. ساختار مگنتایت به گروه مواد سخت معدنی با ساختار $AB_2O_4$ تعلق دارد. شناسایی مگنتایت موجود در فروشاره‌های مغناطیسی از اکسیدهای آهن هموگلوبین موجود در خون کاردشواری بوده و از آنجا که مغناطش اشباع با کاهش اندازه ذرات و اعمال پوشش زیست سازگار بر روی سطح آن‌ها کاهش می‌یابد، بنابراین برای برطرف کردن این مشکل پیشنهاد شده است که از فریت های ترکیبی ($MFe_2O_4$ که M= Co, Zn, Ni, Mn) استفاده شود. از آنجایی‌که ترکیب شیمیایی، مشخصات ریز ساختاری، اندازه و توزیع اندازه ذرات، شکل و مورفولوژی آن‌ها بیشترین تأثیر را روی خواص مغناطیسی دارند، باید روش ساخت مناسبی استفاده نمود که کنترل فوق‌العاده‌ای بر روی این پارامترها داشته باشد[5]. از جمله مهم‌ترین روش‌های شیمیایی‌تر که برای ساخت نانو ذرات مغناطیسی مگنتایت جانشانی شده با دیگر عناصر به کار

رفته، عبارتند از: روش هم رسوبی[1] تخریب حرارتی[2]، میکروامولسیون[3]، پلی‌ال[4]، سولووترمال[5] و سل-ژل[44]. اما روش هیدروترمال به علت تولید محصولاتی با خلوص بالا و استوکیومتری کنترل شده، یکنواختی شکل و مورفولوژی و ریزساختار کنترل شده، بلورینگی بالا و عیوب کم، تولید ذراتی با اندازه‌ی کنترل شده و توزیع اندازه‌ی باریک از طریق تنظیم دمای واکنش، زمان فرآیند، افزودنی‌ها و دیگر فاکتورها، تک مرحله‌ای و دوست‌دار محیط زیست بودن برای سنتز نانوذرات مغناطیسی مورد توجه قرار گرفته است. جانگ[6] و همکارانش تأثیر آلاییدن نانوذرات فریتی با Zn را بر کارایی آن‌ها در افزایش کنتراست MRI و فراگرمایی مغناطیسی مورد قرار دادند. در این پژوهش، ساخت نانوذرات به روش تجزیه گرمایی انجام شد[6]. به علت استفاده از سورفکتانت‌های سمی در سنتز به روش تجزیه حرارتی، استفاده از این روش را برای کاربردهای پزشکی درون تن، دچار مشکل می‌کند[7]. مشکل اساسی استفاده از ذرات در شرایط درون تن، جذب عناصر بیولوژیکی مثل پروتئین‌های پلاسما روی سطح آبگریز ذرات است که باعث می‌شود ذرات سریعا از گردش خون خارج شوند. البته با پوشش آبدوستی که روی سطح ذرات ایجاد می‌کنند این مشکل رفع می‌گردد. اندازه هیدرودینامیکی ذرات در مورد مخفی ماندن آن‌ها از سیستم فاگوسیتوزی بدن نقش مهمی را ایفا می‌کند. هرچه اندازه ذرات بزرگتر باشد زودتر توسط کبد به دام می‌افتد در حالی که ذرات کوچکتر در جریان خون نیمه عمر بیشتری دارند[8]. از این رو زهرایی و همکاران نانوذرات فریت روی با اندازه متوسط 14 نانومتر را در

---

[1] Co-precipitation

[2] High-temperature decomposition of organometallic Precursors

[3] Microemulsion

[4] Polyol

[5] Solvothermal

[6] Jang

حضور سیتریک اسید برای ایجاد پوشش آب دوست بر روی سطح نانوذرات، به روش هیدروترمال، به عنوان عامل کنتراست در تصویربرداری تشدید مغناطیسی، سنتز کردند[9]. در پروژش حاضر سعی می‌شود با تغییر شرایط فرایند هیدروترمال از قبیل میزان عامل احیای سیتریک اسید و pH، ویژگی‌های نانوذرات مغناطیسی مگنتایت جانشانی شده با روی حتی الامکان بهبود یابد.

## مواد و روش تحقیق
### 3-1 مواد اولیه

همه مواد برای ساخت شاره های مغناطیسی شامل نیترات آهن III $Fe(NO_3)_3.9H_2O$، هیدروکسید آمونیوم ۲۵٪ ($NH_4OH$)، نیترات روی $Zn(NO_3)_2$، اسید سیتریک($C_6H_8O_7.H_2O$) و استون از شرکت مرک آلمان با کمینه خلوص ۹۹٪ بود.

### 3-2 ساخت نانو ذرات مغناطیسی مگنتایت جانشانی شده به روش احیای هیدروترمال

برای ساخت نانوذرات مگنتایت جانشانی شده با روی با فرمول $Zn_{0.3}F_{2.7}O_4$، 11/8 میلی مول نیترات آهن III و 1/3 میلی مول نیترات روی در 25 میلی لیتر آب مقطر دی یونیزه حل شد. این محلول توسط یک همزن مغناطیسی به طور مداوم به مدت 20 دقیقه هم زده شد. برای افزایش pH به مقادیر مورد نظر (9، 9/5 و 10) از محلول آمونیوم هیدروکسید استفاده شد. در این شرایط رسوب قهوه‌ای متمایل به نارنجی در ظرف به وجود آمد. برای از بین بردن یون‌های ناخواسته محلول در آب مانند یون‌های آمونیوم و نیترات، عمل شستشوی دقیق رسوبات با سانتریفیوژ انجام شد. در نهایت برای بررسی تأثیر یون‌های سیترات مقادیر 6، 6/5 و 7 میلی مول سیتریک اسید به این رسوبات اضافه شد و در پوشش تفلونی به حجم 320 میلی لیتر ریخته شد. برای انجام فرایند هیدروترمال، پوشش تفلونی درون محفظه فولادی قرار داده شد و در دمای 180 درجه سانتی گراد و به مدت 15 ساعت نگهداری شد. رسوبات سیاه به‌دست آمده چندین بار با استون شستشو داده شد. سپس رسوبات در آون با دمای 70 درجه به مدت 4 ساعت خشک گردید. جدول1. نمونه‌های ساخته شده در مقادیر مختلف از pH و سیتریک اسید

| کد نمونه | دما (درجه سانتی‌گراد) | زمان (ساعت) | عامل احیا (میلی مول) | pH بعد از اضافه کردن آمونیوم هیدروکسید |
|---|---|---|---|---|
| Br.1 | 180 | 15 | 6 | 9 |
| Br.2 | 180 | 15 | 6/5 | 9 |
| Br.3 | 180 | 15 | 7 | 9 |
| Br.3 | 180 | 15 | 7 | 9 |
| Br.4 | 180 | 15 | 7 | 9/5 |
| Br.5 | 180 | 15 | 7 | 10 |

**3-3 مشخصه یابی نانوذرات مگنتایت جانشانی شده با روی**

**3-3-1 بررسی الگوی پراش پرتوی ایکس**

برای تعیین فازهای تشکیل شده آزمایش پراش پرتوی ایکس(XRD) توسط یک دستگاه بروکر مدل D8 ADVANCED با تابش CuKα با طول موج 1/5406 آنگستروم در دانشگاه صنعتی اصفهان انجام شد. برای محاسبه میانگین اندازه بلورک‌ها[1] معادله شرر اصلاح شده به کار گرفته شد.

$$d = 0.9\lambda/\beta cos\theta$$

که در آن λ طول موج پرتوی ایکس، β پهنای قله مورد نظر در نصف شدت بیشینه و θ زاویه براگ قله مورد نظر است.

**3-3-2 میکروسکوپ الکترونی روبشی[2]**

اندازه و توزیع اندازه نانوذرات با استفاده از یک دستگاه میکروسکوپ الکترونی روبشی (SEM) Philips, XL30 در دانشگاه صنعتی اصفهان مورد بررسی قرار گرفت.

**نتایج و بحث**

**1. تأثیر یون‌های سیترات بر روی نانوذرات مگنتایت جانشانی شده با روی $Zn_{0.3}Fe_{2.7}O_4$**

الگوی پراش پرتوی ایکس نمونه‌های ساخته شده در مقادیر مختلف از عامل احیای سیتریک اسید را در زیر مشاهده می‌کنید.

---

[1] Crystallite Size

[2] Scanning electron microscopy

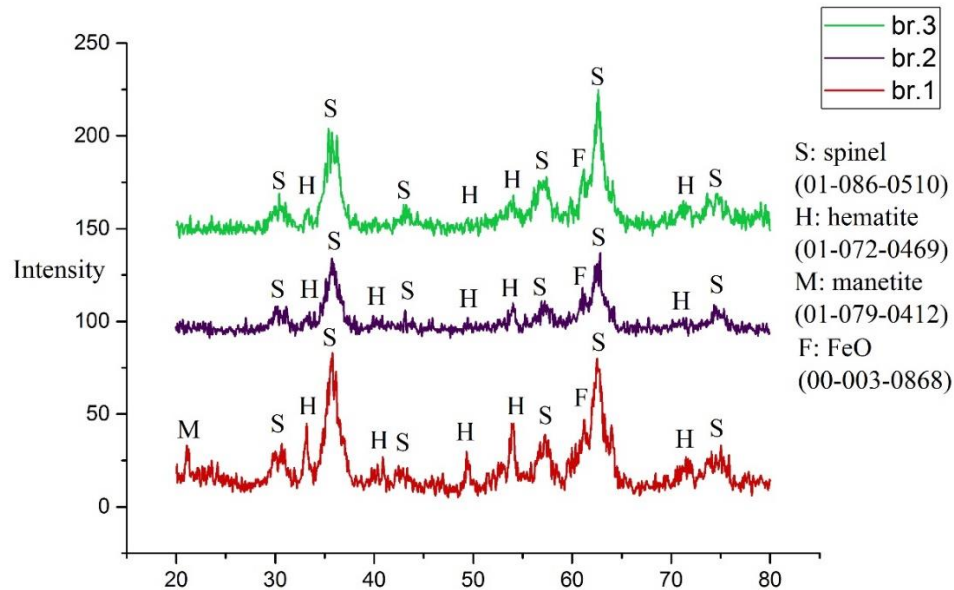

شکل1. الگوی پراش پرتوی ایکس نمونه‌های br.1، br.2 و br.3 در مقادیر مختلف از سیتریک اسید

با توجه به الگوی پراش پرتوی ایکس(XRD) نمونه‌ها در مقادیر مختلف از عامل احیاء، می‌توان نتیجه گرفت با افزایش حضور یون‌های سیترات در محیط واکنش هیدروترمال، میزان تشکیل فاز ناخالص و غیر مغناطیس هماتیت ($\alpha-Fe2O3$) و FeO و اندازه بلورک‌ها همان‌طورکه در جدول 2 مشاهده می‌کنید و از تغییر پهنای قله های پراش یافته مشخص است، کاهش می‌یابد. دلیل آن است که در روش هیدروترمال در حضور یون‌های سیترات، ساز و کار تشکیل مگنتایت انحلال-بازرسوبی خواهد بود به این صورت که پس از ایجاد یون‌های $Fe^{2+}$ در سطح ذرات فری هیدرات، این یون‌ها به دلیل ناپایدارتر بودن پیوند $Fe^{2+}$—O از پیوند $Fe^{3+}$—O از سطح فری هیدرات جدا می‌شوند. حضور یون‌های آهن دوظرفیتی محلول در محیط واکنش، کاتالیستی برای انحلال بیشتر فری هیدرات و آزادسازی یون‌های آهن سه ظرفیتی از سطح فری هیدرات است[10]. حضور یون‌های آهن دوظرفیتی و سه ظرفیتی به طور آزاد در محیط باعث بازرسوبی مگنتایت در محیط واکنش می‌شود[11]. در مقادیر

کمتر از عامل احیای سیتریک اسید بیشتر پیک‌های پراش یافته متعلق به فاز ناخالص هماتیت می‌باشد زیرا در حضور یون‌های سیترات در محیط واکنش، یون‌های آهن III ($Fe^{3+}$) به ($Fe^{2+}$) احیاء می‌شوند. وقتی میزان یون‌های سیترات کم است، مقدار کمتری از آهن III ($Fe^{3+}$) به ($Fe^{2+}$) تبدیل می‌شود در نتیجه میزان فاز اسپینل خالص کاهش می‌یابد و باقی مانده یون‌های ($Fe^{3+}$) به هماتیت تبدیل می‌شوند. با افزایش میزان سیتریک اسید به 7 میلی مول میزان ناخالصی در ساختار کاهش خواهد یافت، و اندازه بلورک‌ها ریزتر خواهد شد که دلیل آن افزایش میزان یون‌های سیترات در محیط واکنش است که از رشد بلورک‌ها جلوگیری خواهد کرد.[12]

جدول 2. میانگین اندازه بلورک‌های نمونه‌های با مقادیر اسید سیتریک مختلف

| کد نمونه | فاز اسپینل(%) | فاز هماتیت(%) | اندازه بلورک اسپینل (nm) |
|---|---|---|---|
| Br.1 | 41/5 | 46/17 | 17/8 |
| Br.2 | 60 | 33/17 | 17/4 |
| Br.3 | 72/25 | 21/75 | 15/44 |

**2. تأثیر pH محلول بر روی ساختار $Zn_{0.3}Fe_{2.7}O_4$**

الگوی پراش پرتو ایکس نمونه‌ها با مقادیر گوناگون pH (9، 9/5 و 10) در شکل 2 نشان داده شده است. میانگین اندازه بلورک‌ها در جدول 3 آورده شده است. مشاهده می‌شود با افزایش pH علاوه بر کاهش بیشتر میزان ناخالصی‌های موجود، بلورینگی ذرات و اندازه بلورک‌ها افزایش می‌یابد. این موضوع نشان می‌دهد که با افزایش pH نرخ جوانه زنی رسوبات فری هیدرات افزایش یافته و ذرات کوچک از بین رفته و ذرات بزرگتر همچنان به رشد خود ادامه می‌دهند همانطور که صفی و همکاران در سنتز نانوذرات فریت کبالت به این نتیجه رسیده‌اند[13].

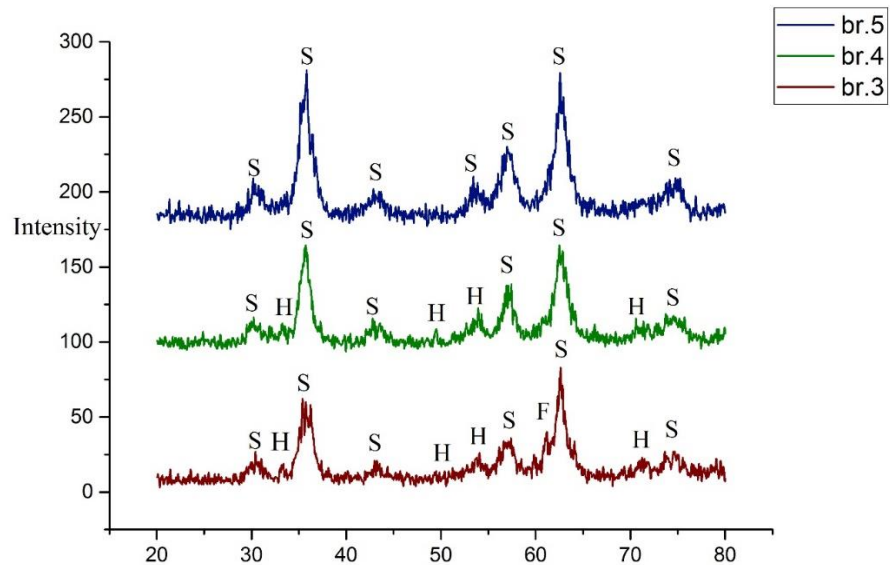

شکل2. الگوهای پراش پرتو ایکس نمونه‌های br3، br4 و br5 در مقادیر مختلف از pH.

جدول 3. میانگین اندازه بلورک‌های نمونه‌های با مقادیر مختلف pH.

| اندازه بلورک اسپینل (nm) | فاز هماتیت(%) | فاز اسپینل(%) | نمونه |
|---|---|---|---|
| 15/44 | 21/75 | 72/25 | Br.3 |
| 16/14 | 17/75 | 82/25 | Br.4 |
| 28 | 0 | 100 | Br.5 |

تصویر میکروسکوپ الکترونی روبشی نمونه br.5 نشان می‌دهد که نانوذرات مگنتایت جانشانی شده با روی به روش هیدروترمال به صورت کروی و یکنواخت در حضور سیتریک اسید به عنوان عامل احیاء سنتز شدند.

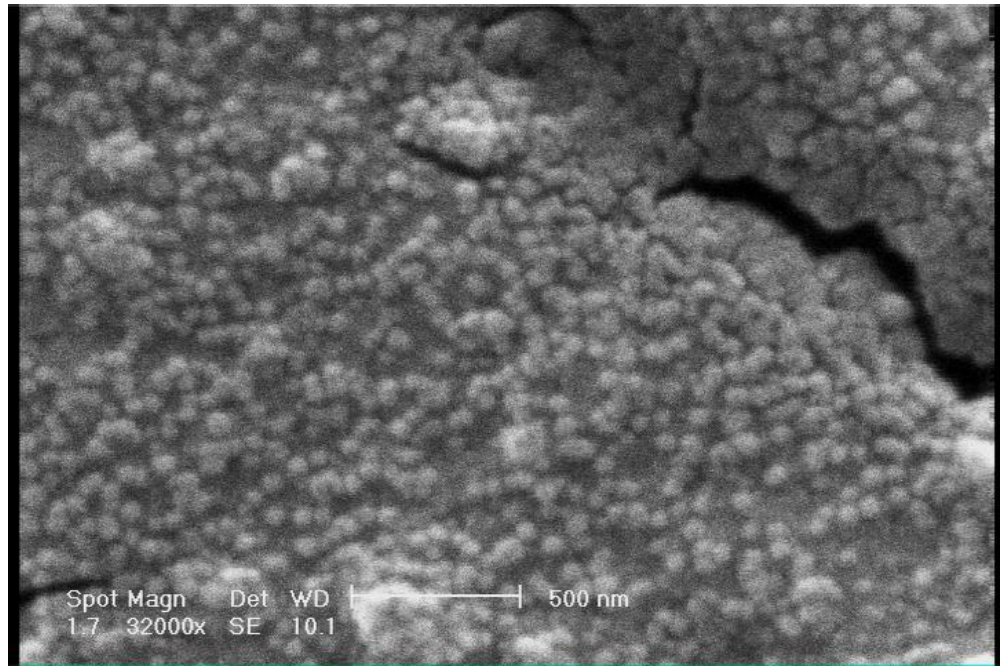

شکل 4-3. تصویر میگروسکوپ الکترونی روبشی (SEM) نمونه br.5

**نتیجه گیری**

نانوذرات مگنتایت جانشانی شده با روی با استوکیومتری $Zn_{0.3}Fe_{2.7}O_4$ به روش هیدروترمال در حضور سیتریک اسید به عنوان عامل احیاء در دمای 185°C به مدت 15 ساعت در مقادیر مختلف از pH و عامل احیاء سنتز شدند. نتایج نشان می‌دهد با افزایش میزان سیتریک اسید مقادیر فازهای ناخالص موجود از قبیل هماتیت کاهش می‌یابد و در pH=10 به فاز اسپینل خالص با اندازه بلورک‌های 28 نانومتر خواهیم رسید.

**مراجع**